\documentclass[12pt,preprint]{aastex}
\usepackage{times,pslatex}
\usepackage{amsmath,amsfonts,graphicx}

\newcommand{\Msun}{\mbox{$M_\odot$}}

\slugcomment{Submitted for publication in the Oldtown Journal of Evil Omens}

\received{2013 Apr 1}
\begin{document}

\title{\it ``Winter is coming''}
\shorttitle{}

\author{
Veselin~Kostov\altaffilmark{1,2},              
Daniel~Allan\altaffilmark{3},
Nikolaus~Hartman\altaffilmark{4,5},
Scott~Guzewich\altaffilmark{6,7},
Justin~Rogers\altaffilmark{8,9}
}


\altaffiltext{1}{Pyke, The Iron Islands, Westeros}
\altaffiltext{2}{Visiting fellow, Moat Cailin, Westeros}
\altaffiltext{3}{Sunspear, Westeros}
\altaffiltext{4}{To whom correspondence should be addressed: nik.hartman@gmail.com}
\altaffiltext{5}{Vale of Arryn, Westeros}
\altaffiltext{6}{Storm's End, Westeros}
\altaffiltext{7}{Exile, Daenerys Targaryen's Entourage, Essos}
\altaffiltext{8}{King's Landing, Westeros}
\altaffiltext{9}{Casterly Rock, Westeros}

\begin{abstract}
Those that do not sow care little about such mundane things as equinoxes or planting seasons, or even crop rotation for that matter. Wherever and whenever the reavers reave, the mood is always foul and the nights are never warm or pleasant. For the rest of the good folks of Westeros, however, a decent grasp of the long-term weather forecast is a necessity. Many a maester have tried to play the Game of Weather Patterns and foretell when to plant those last turnip seeds, hoping for a few more years of balmy respite [1]. Tried and failed. For other than the somewhat vague (if not outright meaningless) omens of {\it ``Winter is Coming''}, their meteorological efforts have been worse than useless. To right that appalling wrong, here we attempt to explain the apparently erratic seasonal changes in the world of G.R.R.M. A natural explanation for such phenomena is the unique behavior of a circumbinary planet. Thus, by speculating that the planet under scrutiny is orbiting a pair of Solar-type stars, we utilize the power of numerical three-body dynamics to predict that, unfortunately, it is not possible to predict either the length, or the severity of any coming winter. We conclude that, alas, the Maesters were right -- one can only throw their hands in the air in frustration and, defeated by non-analytic solutions, mumble {\it ``Coming winter? May be long and nasty ($\sim 850$ days, $T<268K$) or may be short and sweet ($\sim 600$ days, $T\sim273K$). Who knows...''}.
\end{abstract}
\include{variables}
\section{Introduction}
\label{sec:intro}

Winters, while fun and exciting during the holiday season and good business for struggling retail stores, are rather miserable. Imagine how much worse it would be if winter's length and severity varied capriciously. Enter the notoriously unpredictable seasons promoted by one G.R.R.M. [2]. Here we attempt a long-term forecast for the unusual weather endured by the inhabitants of Westeros and Essos, for the benefit of their loyal citizens and mighty lords.

To begin, we give audience to simple explanations for the unusual climes in \emph{A Song of Ice and Fire}. We scrutinize them and then consign their fates to the judgements of the Old Gods. Those that do not pass will be needed at the Wall.

\subsection{Axial Tilt}
\label{sec:tilt}

The stable tilt of Earth's axis keeps us well out of all this trouble. Although some have blamed Westerosi seasons on a wobbly planetary axis, we reject this suggestion as nothing more than sinister Casterly Rock propaganda. As even Ser Gregor could tell you, a planetary moon [3, 4] precludes axial excursions by stabilizing the tilt. 

\subsection{Eccentric Orbit}
\label{sec:ecc}

Myriad extrasolar planets, discovered over the past two decades\footnote {http://exoplanet.eu/catalog/}, have undermined our solar system's sense of importance and expanded the variety of planetary orbits. Consider the cruel Eccentric Orbit, which exacts harsh, long winters while its planet dallies far from its star. Is this the planet we are looking for? The winters fit, but, no, we have gone too far. The privileged might stay warm [5], but suffering farmers, rushed to sow and reap during a ruthlessly short perihelion, would ultimately embrace the Cold Side and join the swelling ranks of the Others [6]. The continents would be overrun by lifeless husks. 

\subsection{Pollutants}
\label{sec:geo}

To the best of our knowledge, industrial emissions are not to blame. But before we dismiss the issue, we must indict that most terrifying of greenhouse gas producers, whose carbon footprint is dwarfed only by its actual footprint. We defer further discussion on dragons [7] and all matters geophysical for a later work (in prep.).

\subsection{Magic}
\label{sec:magic}

Yep, this should work. 

\subsection{Other (no ``s''!)}
\label{sec:other}

Finally, all things said and done, the only reasonable explanation remains in the arcane physics of three-body systems, where unpredictability and chaotic behavior is the name of the Game. Which is, naturally, played out in the next section.

This paper is organized as follows. In Section \ref{sec:main} we describe our analysis, followed by discussion of the results in Section \ref{sec:disc}. We draw our conclusion in Section \ref{sec:end}.
\section{Materials and Methods}
\label{sec:main}

One sun and one planet give periodic, vanilla orbits. We must abandon Keplerian monogamy and explore multiple bodies. Fortunately, the many-body problem has been solved repeatedly (some might say gratuitously) in Westeros. Our planet will have two stars.

We simulate the orbit of the circumbinary planet (CBP) around a binary star composed of two Solar-type stars. We use a numerical three-body integrator in hierarchical Jacobian coordinates to calculate the position and velocity of the CBP at each point along its orbit. The integrator starts when two stars and the planet are aligned along the same axis. It may be a choice of convenience, but we suspect it is portentous. The parameters of the system are outlined in Table \ref{tab:param}. The planet is dangerously close (but outside) the critical distance for dynamical stability. For simplicity, we assume the three bodies are coplanar, the mass of the planet is negligible and its axis is not tilted. At each time step, we compute the distance from the planet to the two stars and, using the black-body approximation, calculate the equilibrium surface temperature of the planet by: 

\begin{equation}
T_p = [\frac{1}{4}(1-A_B)]^{1/4}[(\frac{R_1}{r_1})^2 T_1^4 + (\frac{R_2}{r_2})^2T_2^4]^{1/4}
\label{eq:eq1}
\end{equation} 

where $T_{p}$ is the equilibrium surface temperature of the planet, $A_B$ is Bond albedo of $0.3$ and greenhouse warming due to presence atmosphere of 30 degrees (see clouds, dragon smoke, Section \ref{sec:geo}), $R_1$ and $R_2$ are the two stellar radii (each equal to $1.R_{Sun}$), $T_1$ and $T_2$ are the two stellar effective temperatures (both equal to $6000K$) and $r_1$ and $r_2$ are the two instantaneous distances from the planet to each of the two stars. 

\begin{figure}
\epsscale{1.0}
\plotone{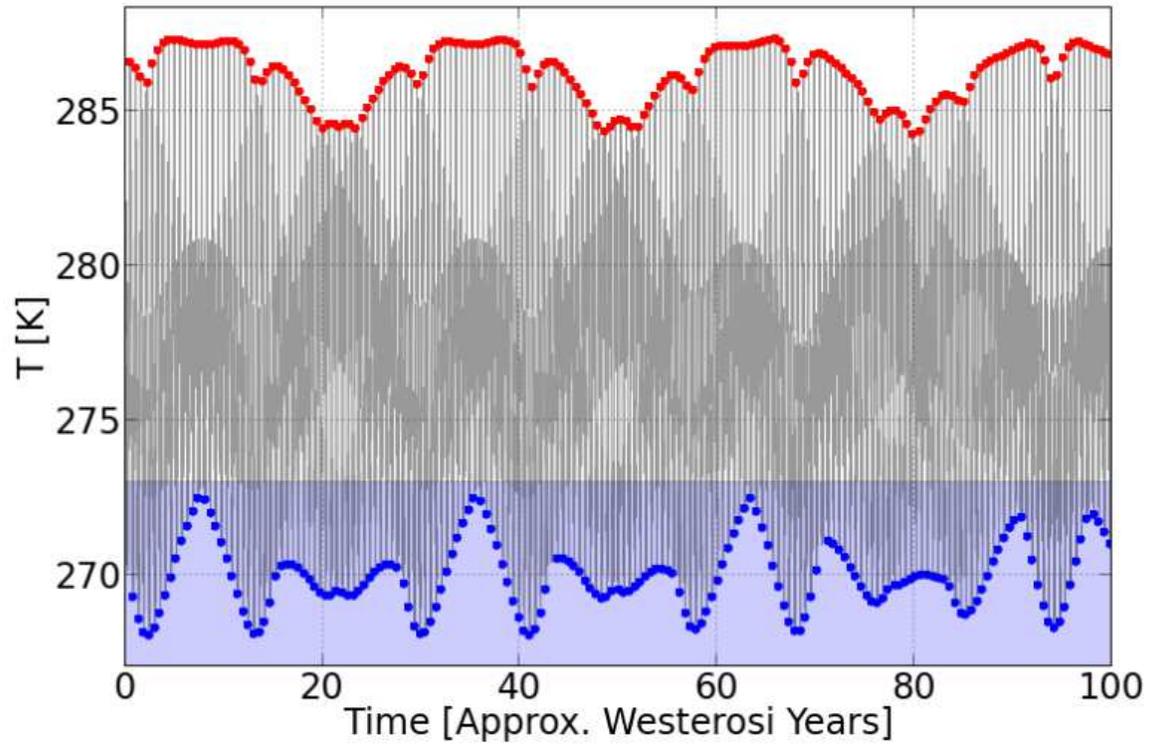}
\centering
\caption{Surface temperature of a circumbinary planet orbiting two 6000K, $1\Msun$ stars on a $\sim 750$-day orbit. Blue symbols indicate individual winters, red symbols -- individual summers. Everything inside the light blue region is frozen solid. 
\label{fig:fig1}}
\end{figure}

The surface temperature as a function of time is shown on Figure \ref{fig:fig1}. The seasonal cycle varies significantly, due to the continuously changing semi-major axis of the planet and corresponding variations of surface insolation. The distance of the planet to the barycenter of the system can be as low as $\sim1.5$ AU and reaches up to $\sim2.5$ AU.

To study the long-term stability of the CBP, we embarked on a long voyage of integration. The orbit was stable after integrating for one million days. On Day 1,000,001 the Andals invaded. Funding cuts to all research projects promptly followed. We were unable to continue our work. Those of us who expressed strong opinions against the cuts were sent to the Wall. 

\section{Discussion}
\label{sec:disc}

The complicated seasonal pattern for $\sim100$ Westerosi years shown in Figure \ref{fig:fig1} is striking. Both the severity (defined here as many consecutive days below freezing point) and duration of winters vary eratically, with uncomfortably wild swings. Average temperature differences between winters can vary drastically, by as much as 5 degrees; summers behave in a similar fashion, with an amplitude of $\sim2$ degrees. We remind the reader that a typical Westerosi year is $\sim700$ Earth days. We note that a non-zero planetary obliquity or the presence of additional planets in the system may significantly modify our results.

While there is an overall and somewhat regular ``moving envelope'' of multi-year spells of (hot summers+cold winters) followed by (cool summers+mild winters), neither the average length of these periods, nor their detailed, long-term behavior can be known a-priori. It may be tempting to suggest that mild summers typically correspond to balmy winters [8] but this is not necessary be the norm. A sequence of balmy winters and chilly summers may or may not be followed by a set of particularly unpleasant winters (cf. Fig. \ref{fig:fig1}, near years 28, 85). An emerging pattern, and one that may save many a crop, is that local minima in the upper envelope of summer temperatures (cf. Fig. \ref{fig:fig1}, near years 30, 40, 93) can be interpreted as the harbingers of terrible winters.

\begin{figure}
\centering
\plotone{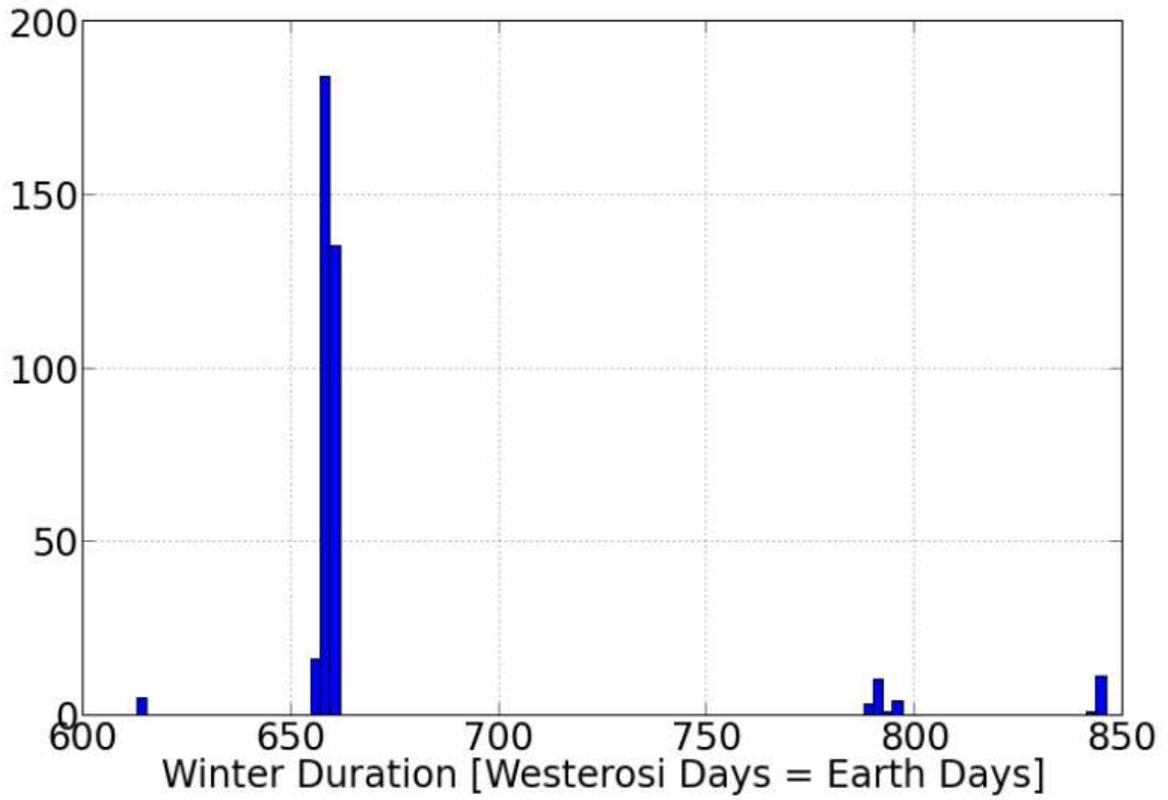}
\caption{A histogram of winter durations. Shortest winters can last for as little as $\sim600$ Westerosi days while longest can be up to 850 days.
\label{fig:fig2}}
\end{figure}

Another very intriguing result is the distribution of the duration of winters. In Figure \ref{fig:fig2} we show a histogram of all the points from the lower envelope (blue symbols) in Fig. \ref{fig:fig1}. The distribution is strongly peaked around the typical orbital period. However, there are a number of outliers on either side of the average value. Some winters can be as short as $\sim600$ Westerosi days (=Earth days), while others as long as $\sim850$!

\section{Conclusion}
\label{sec:end}

With heavy hearts, we conclude that our attempts to provide the good folks of Westeros with a reliable weather forecast are inconclusive. 
\section{Acknowledgements}

The authors gratefully acknowledge everyone who has supported this crazy idea. We are especially hopeful our advisors never see this work as we would be hard pressed to answer the logical question {\it "...Why are you wasting your time? The thesis won't write itself, Mister!..."}. 

This research used our allegedly free time, an obscene amount of coffee and some mediocre brainpower.
\section{References}

[1] {\it ``...Tyrion shut the heavy leather-bound cover on the book he was reading, a hundred-year-old discourse on the changing of the seasons by a long-dead maester...''}, Chapter {\it Tyrion},  Game of Thrones

[2] {\it ``...They say it grows so cold up here in winter that a manÕs laughter freezes in his throat and chokes him to death...''}, Chapter {\it Eddard},  Game of Thrones

[3] {\it ``...Aeron had seen his eldest brother not a moonÕs turn past...''}, Chapter {\it The Prophet},  A Feast for Crows

[4] {\it ``...SheÕs mad with moon blood...''}, Chapter {\it Brienne},  A Feast for Crows

[5] {\it ``...In winters past, food could be brought up the kingsroad from the south, but with the war...''}, Chapter {\it Jon},  A Dance with Dragons

[6] {\it ``...The Others. They are mentioned in the annals, though not as often as I would have thought...''}, Chapter {\it Jon},  A Dance with Dragons

[7] {\it ``...Viserion hissed again. Smoke rose between his teeth, and deep down in his throat they could see gold fire churning...''}, Chapter {\it Daenerys},  A Dance with Dragons

[8] {\it ``...It was the year of false spring, and he was eighteen again, down from the Eyrie to the tourney at Harrenhal...''}, Chapter {\it Eddard},  A Game of Thrones
\begin{table}[ht]
\begin{center}
\caption{Parameters of the World of Westeros
\label{tab:param}}
\begin{tabular}{l|ll|l}
\hline
\hline
-- & {\bf Binary Star} & & {\bf Circumbinary Planet} \\
-- & Primary & Secondary & \\
\hline
Mass [\Msun] & 1.0 & 1.0 & 0.0 \\
$T_{eff}$ & 6000K & 6000K & see Fig. \ref{fig:fig1} \\
Metalicity & Valyrian Steel & Valyrian Steel & Valyrian Steel \\
Age & Old Nanny's Nanny's & Old Nanny's Nanny's & Old Nanny's \\
\hline
Semimajor Axis [AU] & 0.55 & -- & 2.05 \\
Period [Days] & 110.0 & -- & $\sim700.0$\\
Eccentricity & 0.5 & -- & 0.1 \\
Orbital Inclination [degrees] & 90.0 & -- & 90.0 \\
Albedo & -- & -- & 0.3 \\
\hline
\hline
\end{tabular}
\end{center}
\end{table}
\end{document}